\documentclass[aps,prb,twocolumn,amsmath,amssymb]{revtex4}
\usepackage{color}
\usepackage{verbatim}
\usepackage{hyperref}
\usepackage{epsf}
\usepackage{graphics}
\usepackage{graphicx}
\usepackage{natbib}
\usepackage{dcolumn}
\usepackage{bm}

\begin{document}
\title{Theory of the spin nematic to spin-Peierls quantum phase transition\\ in ultracold spin-1 atoms in optical lattices}
\author{Christoph M. Puetter}
\email{cpuetter@physics.utoronto.ca}
\affiliation{Department of Physics, University of Toronto, Toronto, 
Ontario, Canada M5S 1A7}
\author{Michael J. Lawler}
\affiliation{Department of Physics, University of Toronto, Toronto, 
Ontario, Canada M5S 1A7}
\author{Hae-Young Kee}
\email{hykee@physics.utoronto.ca}
\affiliation{Department of Physics, University of Toronto, Toronto, 
Ontario, Canada M5S 1A7 and \\
School of Physics, Korea Institute for Advanced Study, Seoul 130-722, Republic of Korea}
\date{\today}
\begin{abstract}
We present a theory of the anisotropy tuned quantum phase transition between spin nematic
and spin-Peierls phases in $S=1$ systems with 
significant bi-quadratic exchange 
interactions. Based on quantum Monte Carlo studies on finite size systems, 
it has been proposed that this phase transition 
is second order with new deconfined fractional excitations that are absent in 
either of the two phases. 
The possibility of a weak first order 
transition, however, cannot be ruled out.
To elucidate the nature of 
the transition, we construct a large-$N$ SO(3$N$) model for 
this phase transition and 
find in the $N\to\infty$ limit that the transition is 
generically of first order. 
Furthermore, we
find a critical point in the one-dimensional (1D) limit, where two 
transition lines, separating spin-nematic, ferromagnetic,
and spin-Peierls phases, meet.
Our study indicates that the spin-nematic phase is absent in 1D, 
while its correlation length diverges at the critical point.
Predictions for $^{23}$Na 
atoms trapped in an optical lattice, 
where the nematic to spin-Peierls quantum phase 
transition naturally arises, are discussed.
\end{abstract}

\maketitle

\section{Introduction} 
Conventional Landau-Ginzburg theory describes
continuous phase transitions by fluctuations of an order 
parameter. 
A nonzero value of the order parameter signals  
spontaneous symmetry breaking and the presence of an ordered phase. 
Moreover, if several distinct ordered ground states are possible,
Landau-Ginzburg theory generically predicts
that ordered phases with unrelated broken symmetries are separated by 
either intervening phases or a first order phase boundary.
In the simplest case, this is illustrated by the free energy 
for two independent order parameters $\phi_{1}$ and $\phi_{2}$ as 
\cite{Chaikin05Book}
\begin{equation}
  f = \frac{r}{2} (\phi_{1}^{2} + \phi_{2}^{2}) 
  - \frac{g}{2} (\phi_{1}^{2} - \phi_{2}^{2}) 
  + u_{1} \phi_{1}^{4} +  u_{2} \phi_{2}^{4} 
  + 2 u_{12} \phi_{1}^{2} \phi_{2}^{2}.
\end{equation}
Here, a direct second-order transition 
in the $r$-$g$ parameter space
between phases with either nonzero $\phi_{1}$ or $\phi_{2}$ 
requires fine tuning 
such that $u_{1} u_{2} = u^{2}_{12}$ is satisfied.

Recently, a theory of critical phase transitions has been proposed 
that reaches beyond the Landau-Ginzburg paradigm and allows
direct continuous transitions between distinct broken-symmetry phases.
\cite{Senthil04Science,Senthil04PRB}
In this theory, the relevant degrees of freedom are described in terms of 
fields that carry fractional quantum numbers 
and become deconfined only at the critical point.
In particular, it was argued that a direct continuous transition
from a valence bond solid (VBS) to a N\'{e}el phase 
\cite{Senthil04Science,Senthil04PRB}
or a spin-nematic phase
\cite{Harada07JPSJ,Grover07PRL} 
falls into this class of phase transitions.
For this reason the search for  
such deconfined critical phenomena (DCP) 
in simple model systems has attracted much attention.

Several systems have been studied up to now.
Based on quantum Monte Carlo simulations
on the two-dimensional spin-1/2 Heisenberg model with
an additional four-spin interaction (``$JQ$ model'')
the phase transition between a N\'{e}el ordered state and a 
VBS appears to be consistent with the deconfined critical scenario.
\cite{Sandvik07PRL,Melko08PRL}
This, however, has been disputed by  
other numerical studies 
\cite{Kuklov08CondMat,Jiang07CondMat,Kuklov06AnnPhys}
that support a weakly first order transition.
Another candidate system is the $Sp(4)$-Heisenberg model on a 
square lattice, which describes spin-3/2 cold atom systems
and has also been conjectured to harbor a direct second-order transition
between a N\'{e}el and a VBS state. 
\cite{Qi08CondMat} 

The present approach focuses on the 
anisotropic bilinear bi-quadratic spin-1 Heisenberg model 
(``$JK$ model'') on a square lattice
\begin{eqnarray}
  \label{eq:JKModel}
  H &=& \sum_{i}
  \big[ J {\bf S}_{i} \cdot {\bf S}_{i + x}
    + K \big({\bf S}_{i} \cdot {\bf S}_{i + x}\big)^{2}\big] \nonumber \\
   & &+ \lambda \sum_{i}
  \big[ J {\bf S}_{i} \cdot {\bf S}_{i + y}
    + K \big({\bf S}_{i} \cdot {\bf S}_{i + y}\big)^{2}\big]  
\end{eqnarray}
where the exchange integrals for neighboring spins 
in $y$ direction are reduced by an anisotropy  parameter $\lambda$ 
(i.e., $J \rightarrow \lambda J$ and $K \rightarrow \lambda K$ on $y$ bonds).
For $\lambda = 0$, this Hamiltonian describes decoupled spin chains, while for
$\lambda = 1$ the full square lattice symmetry is recovered.   
The parameter range of interest is specified by
$K < J < 0$, which is also thought
to be the natural range for $^{23}\text{Na}$ atoms 
in an optical lattice. \cite{Yip03PRL,Imambekov03PRA} 
In this range the quadratic term favors ferromagnetic order, while the 
quartic term prefers the formation of singlet bonds.
This competition is solved in two dimensions ($\lambda=1$) by having a
spin-nematic ground state 
(which breaks spin-rotational symmetry but preserves
time-reversal symmetry $\langle {\bf S}_{i} \rangle$ = 0) 
and in one dimension ($\lambda = 0$) by forming a dimerized 
ground state (which breaks translational invariance). 
\cite{Imambekov03PRA,Yip03PRL,Harada07JPSJ,Laeuchli06PRB,Lewenstein08AdvPhys,Podolsky05NJP}
When the anisotropy parameter is continuously changed between 1 and 0,
quantum Monte Carlo studies by Harada et~al. \cite{Harada07JPSJ} 
suggest that the system undergoes a 
Landau-forbidden direct second-order transition between the nematic
and the dimer phases, which further motivated  
a recently developed continuum theory 
for the nematic-dimer phase transition based on DCP.
\cite{Grover07PRL} 
However,  because of significant finite-size effects, they
were unable to rule out a weak first-order transition or the existence
of two successive phase transitions. \cite{Harada07JPSJ} 

In this paper, we examine the nematic-dimer phase transition 
in anisotropic spin-1 systems using two complementary approaches.
The first approach is based on the bond operator formalism
introduced by Chubukov \cite{Chubukov90PRB} 
and is particularly suited for studying 
the dimer phase as pairs of neighboring spins are described
by common bosonic bond operators.
Taking the classical limit of this approach by neglecting
all quantum fluctuations, one can already 
obtain a good overview over the $JK$ model.
Complementary to the bond operator method,
we then construct an SO($3N$) model 
and study its large-$N$ limit ($N=1$ is the physical limit).
This approach explicitly 
takes the disordering effects of zero-point fluctuations into account.
For $N \rightarrow \infty$ we 
find that a first-order phase boundary separates the spin nematic
from gapped spin-liquid phases except at special SU($3N$) 
symmetric points, where the transition becomes second order.
In the 1D ($\lambda = 0$) limit, the existence of the spin-nematic phase 
has been studied extensively but remains elusive.
\cite{Chubukov90PRB,Buchta05PRB,Rizzi05PRL,Rossini05JPhysB}
In this limit, we find a critical point at $J=K<0$ where 
spin-nematic, ferromagnetic and spin-Peierls correlation functions
diverge.
The spin nematic phase, however, does not exist in a finite 
parameter range near $J=K<0$ at $\lambda=0$.
Finally, predictions for $^{23}$Na atoms in an 
optical lattice, where the nematic to dimer quantum phase
transition naturally arises, are presented.  

\section{Spin-$1$ bond operator model}
To gain a simple understanding of the anisotropic $JK$ model of Eq. 
\eqref{eq:JKModel}, consider the classical limit of the bond operator 
model following Chubukov \cite{Chubukov90PRB}. Grouping the $N_s$ sites
of the square lattice into $N_s/2$ bonds in a columnar pattern, we may 
reformulate the $JK$ model in terms of bosonic operators on 
these bonds that create and 
annihilate singlets  $|00\rangle = \hat s^\dagger|0\rangle$, triplets 
$|1,m_t\rangle = \hat t_{m_t}^\dagger|0\rangle$ (where $m_{t} = 0, \pm 1$),
and quintuplets $|2,m_q\rangle = \hat q^\dagger_{m_q}|0\rangle$ 
(where $m_{q} = 0, \pm 1, \pm 2$). A constraint of one boson per site is 
then necessary to stay within the physical Hilbert space. Allowing
the bosons to completely condense (as in Bogoliubov theory) then produces a 
simple phase diagram based on the energy of different possible condensates.

To carry out such program, we must first re-construct the spin Hamiltonian of Eq. \eqref{eq:JKModel} in terms of the above-introduced bosons. On a given bond $\ell=(i,i+\hat x)$, it is easier to work with the generators $\vec L_\ell=\vec S_i+\vec S_{i+\hat x}$ and $\vec M_\ell = \vec S_i - \vec S_{i+\hat x}$ than with the spin operators directly. Direct evaluation of the matrix elements of these operators in the total spin basis of bond $\ell$ then tells us how to represent them in terms of singlet, triplet and quintuplet operators. While a little tedious, the net result is (dropping the bond labeling $\ell$):
\begin{align}
\hat L^z &= \sum_{m_{t}} m_t t^\dagger_{m_{t}}t_{m_{t}} +
    \sum_{m_{q}} m_qq^\dagger_{m_{q}}q_{m_{q}}
\end{align}
\begin{align}
\hat L^+ &= \sqrt{2}\big(t^\dagger_1t_0 + t^\dagger_0t_{-1}\big) + 
   2\big(q^\dagger_2q_1 + q^\dagger_{-1}q_{-2}\big) \\\notag&\ \ \ \ + 
   \sqrt{6}\big(q^\dagger_1q_0 + q^\dagger_0q_{-1}\big)
\end{align}
\begin{align}
\hat L^- &=\sqrt{2}\big(t^\dagger_0t_1 + t^\dagger_{-1}t_{0}\big) + 
   2\big(q^\dagger_1q_2 + q^\dagger_{-2}q_{-1}\big) \\\notag&\ \ \ \ + 
   \sqrt{6}\big(q^\dagger_0q_1 + q^\dagger_{-1}q_{0}\big)
\end{align} 
\begin{align}
 \hat M^z &= \sqrt{\frac{8}{3}}\big(t^\dagger_0 s + s^\dagger t_0\big) +
     \sqrt{\frac{4}{3}}\big(q^\dagger_0t_0 + t^\dagger_0q_0\big) \\\notag&\ \ \ \  +
     q^\dagger_1t_1 + t^\dagger_1q_1 + q^\dagger_{-1}t_{-1}+t^\dagger_{-1}q_{-1}
\end{align} 
\begin{align}
\hat  M^+ &=-\frac{4}{\sqrt{3}}\big(t^\dagger_1 s -s^\dagger t_{-1}\big) 
    -2\big( q^\dagger_2t_1 - t^\dagger_{-1}q_{-2}\big)\\\notag&\ \ \ \  
    -\sqrt{2}\big(q^\dagger_1 t_0 - t^\dagger_0 q_{-1}\big)
    -\sqrt{\frac{2}{3}}\big(q^\dagger_0 t_{-1} - t^\dagger_1 q_0\big)
\end{align}
\begin{align}
 \hat M^- &=  -\frac{4}{\sqrt{3}}\big(t^\dagger_1 s -s^\dagger t_{-1}\big) 
    -2\big( q^\dagger_2t_1 - t^\dagger_{-1}q_{-2}\big) \\\notag&\ \ \ \  
    -\sqrt{2}\big(q^\dagger_1 t_0 - t^\dagger_0 q_{-1}\big)
    -\sqrt{\frac{2}{3}}\big(q^\dagger_0 t_{-1} - t^\dagger_1 q_0\big) 
\end{align}

To compute the condensate energies for various phases, it is then useful to group the bosonic operators into a single vector $\vec\psi=(s,t_1,t_0,t_{-1},q_2,...,q_{-2})$ and to express the above operators in the compact form $\hat L^z = \psi^\dagger_{\alpha}L^z_{\alpha,\beta}\psi_{\beta}$, etc. Substituting these expressions into the Hamiltonian then leaves us with the generic form
\begin{equation}
  H = \sum_{\ell} h^{(1)}_\alpha \psi^\dagger_{\ell, \alpha} \psi_{\ell, \alpha} + 
  \sum_{\langle \ell\ell'\rangle} h^{(2)}_{\alpha,\beta,\gamma,\delta}(\ell,\ell')
  \psi^{\dagger}_{\ell,\alpha}\psi^\dagger_{\ell',\beta}\psi_{\ell',\gamma}\psi_{\ell,\delta}
\end{equation}
where $h^{(1)}$ and $h^{(2)}$ depend on $J/K$. That this Hamiltonian consists of only one-body and two-body terms results from the one boson per bond constraint. Assuming the bosons condense $\psi_{\ell,\alpha} = \langle\psi_{\ell,\alpha}\rangle + \delta\psi_{\ell,\alpha}$, we then extract the leading contribution to the condensate energy by expanding the ground-state energy to zeroth order in powers of $\delta\psi_{\ell,\alpha}$.

Four phases are of particular interest: a ferromagnetic phase with all spins pointing up,
a dimerized spin-Peierls phase with a singlet on each bond, a ``spin nematic'' phase 
in which each site is in the state $|1,0\rangle$ and an antiferromagnetic phase with
$|1,1\rangle$  ($|1,-1\rangle$) on the A (B) sublattice. The bosonic condensates
for these idealized states are:  $\langle q_2\rangle=1$ in the ferromagnetic phase; 
$\langle s\rangle=1$ in the dimer phase; $\langle q_0\rangle = \sqrt{2/3}$ and 
$\langle s\rangle = 1/\sqrt{3}$ in the spin-nematic phase; and $\langle s\rangle = 1/\sqrt{3}$, 
$\langle t_0\rangle = (-1)^y/\sqrt{2}$ and $\langle q_0\rangle=1/\sqrt{6}$ in the antiferromagnetic
phase (the sign here alternates along a column in the columnar dimer pattern). 

\begin{figure}[t]
  \includegraphics*[angle=0, width=0.9\linewidth, clip]{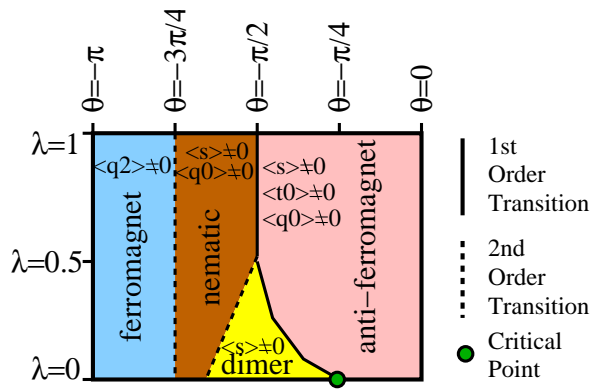}
  \caption{(Color online) Classical phase diagram. All phases are described by
  a finite condensate of bosonic bond operators following 
  Chubukov (Ref. \onlinecite{Chubukov90PRB}). 
  Note the existence of a spin-nematic phase in the
  one-dimensional ($\lambda=0$) limit.
    \label{fig:S2pdiagram}}
\end{figure}

Using the above condensates as a guide, we construct a simple phase diagram by minimizing the
condensate energy as a function of $\langle s\rangle$, $\langle t_{m_t}\rangle$, and $\langle q_{m_q}\rangle$ on at most two independent neighboring bonds (that is, we explore an 18-variational-parameter space). This approach follows that of Ref. \onlinecite{Chubukov90PRB} in essence and can be thought of as a simple two-site clustering method. The resulting phase diagram is depicted in Fig. \ref{fig:S2pdiagram}. All phases except the ferromagnetic phase have a finite dimerization due to the explicit translational symmetry breaking of this approach. 

It is worth noting the finite existence of a spin-nematic phase in the one-dimensional $\lambda=0$ limit as shown in Fig. \ref{fig:S2pdiagram}. This phase is not expected to survive in the presence of quantum fluctuations (finite $\delta\psi_{\ell,\alpha}$) as pointed out by Chubukov \cite{Chubukov90PRB}. 
However, the resulting phase diagram after these effects are included is 
not clear from this approach. 
The absence or presence of a (gapped) spin nematic in the 1D limit
has been the focus of extensive numerical studies
based on density matrix renormalization group, exact diagonalization, quantum state transfer, etc.
\cite{Fath95PRB,Laeuchli06PRB,Rizzi05PRL,Rossini05JPhysB,Romero-Isart07PRA} 
This issue still remains to be settled
as diverging correlation lengths near the ferromagnetic phase 
limit numerical approaches. 
We will show below that 
in the large-$N$ limit of the $JK$ model 
the spin nematic phase does not occupy 
a finite parameter region near the critical point $J/K=1$ for $\lambda = 0$,
while, approaching the critical point, spin-nematic 
correlations certainly diverge.
This suggests that the one-dimensional spin-nematic phase vanishes 
even for $N=1$, although $1/N$ fluctuations may 
qualitatively change the universality class of the phase transition 
from that of the large-$N$ limit.

\section{Spin-$1$ Schwinger boson model}
In contrast to the bond operator approach,  
large-$N$ methods include the disordering effects of quantum fluctuations.
For the present system, one therefore might expect that 
fluctuations will help restore the translational invariance 
in the dimerized nematic phase and affect the nature and the location of
the phase boundary between the nematic and the dimer phase.  
Embarking on the large-$N$ route, we introduce bosonic SU(3) spinors
$a_{i, \alpha}$ ($\alpha = 1, 2, 3$) to  
rewrite the spin operators on each site $i$ as \cite{Zhang05CondMat}
\begin{equation}
  \label{eq:triplons}
  S_{i, \alpha} = -\text{i} 
  \varepsilon_{\alpha \beta \gamma} a^{\dagger}_{i, \beta} a_{i, \gamma}.
\end{equation}
In this representation, $a^{\dagger}_{i, \alpha}$ 
creates a particle in a state whose spin component
in the $\alpha$ direction is zero, that is,
for a  given quantization axis, we define
$a^{\dagger}_{i, 1} |0\rangle 
= (|1,-1\rangle_{i} - |1,1\rangle_{i})/\sqrt{2}$,
$a^{\dagger}_{i, 2} |0\rangle 
= -i (|1,-1\rangle_{i} + |1,1\rangle_{i})/\sqrt{2}$, and
$a^{\dagger}_{i, 3} |0\rangle = |1,0\rangle_{i}$.
Imposing single occupancy on average on each site via a
Lagrange multiplier $\mu$, one finds for the
anisotropic $JK$ model
($\lambda_{x} = 1$, $\lambda_{y} = \lambda$)
\begin{eqnarray}
  \label{eq:JKModeltriplon}
  H &=& \sum_{\stackrel{\rho = x, y}{i}}  \lambda_{\rho}
  \Big[(-J + K)
  a_{i, \alpha}^{\dagger} a_{i + \rho, \alpha}^{\dagger} 
  a_{i, \beta} a_{i + \rho, \beta} \\
  & & + J a_{i, \alpha}^{\dagger} a_{i + \rho, \alpha} 
  a_{i, \beta}^{\dagger} a_{i + \rho, \beta} \Big] 
  - \mu \sum_{i} \big(a_{i, \alpha}^{\dagger} a_{i, \alpha} - 1 \big), 
  \nonumber
\end{eqnarray}
where the summation over the spinor components is implied
and the exchange integrals are commonly parametrized by an angle $\theta$,
i.e., $J=\text{cos}(\theta)$ and $K=\text{sin}(\theta)$.
The first term in this Hamiltonian possesses 
uniform SU(3) symmetry, while the second term has staggered SU(3) symmetry
on a bipartite lattice.
As a result, global SU(3) symmetry is found for special values of $J/K$,
namely, for $\theta = -3\pi/4, -\pi/2, \pi/4$, $\pi/2$.
\cite{Li04NJP}

The order parameter of a spin nematic 
is a symmetric, traceless rank-2 tensor constructed from the  
spins, i.e.,
\begin{equation}
  Q_{\alpha \beta}(i) 
  =  \frac{S_{i, \alpha} S_{i, \beta} + S_{i, \beta} S_{i, \alpha}}{2}
  - \frac{2}{3} \delta_{\alpha \beta}.
\end{equation}
This tensor can be expressed in terms of the triplons $a_{i, \alpha}$,
which when condensed, $\langle a_{i, \alpha} \rangle \neq 0$,
describe long-range order.
Up to an SU(2) rotation, one then finds that condensation  
of one of the spinor components
$\langle a_{i, \alpha} \rangle \propto x_{\alpha}$ 
corresponds to nematic order, and
condensation of two components 
$\langle a_{i, \alpha} \rangle \propto  x_{\alpha}$ 
and $\langle a_{i, \beta} \rangle \propto x_{\beta}$ 
($\alpha \neq \beta$),
where $x^{*}_{\alpha} x_{\beta}$ is purely complex,
indicates ferromagnetic order, 
while additionally staggered expectation values with
$\langle a_{i, \alpha (\beta)} \rangle \propto x_{\alpha (\beta)}$ 
on one sub-lattice ($i \, \epsilon \, A$) and
$\langle a_{i, \alpha (\beta)} \rangle \propto x^{*}_{\alpha (\beta)}$ 
on the other ($i \, \epsilon \, B$)
are associated with anti-ferromagnetic order.

Now, consider generalizing to 
a large-$N$ SO($3N$) model [with SU($3N$) symmetry at 
$\theta=-3\pi/4,-\pi/2$, etc.] and expanding
in powers of $1/N$.
To leading order, we obtain a mean-field theory
with self-consistent equations
\begin{eqnarray} 
  \chi_{\rho} 
  &=& \langle \frac{1}{N} \sum_{\alpha=1..3N} 
  a_{i, \alpha}^{\dagger} a_{i + \rho, \alpha} \rangle \\
  \eta_{\rho} 
  &=& \langle \frac{1}{N} \sum_{\alpha=1..3N} 
  a_{i, \alpha}^{\dagger} a_{i + \rho, \alpha}^{\dagger} \rangle,
\end{eqnarray}
where $\rho = x$ or $y$. 
These fields describe short-range correlations and we choose them 
to be real to ensure an expected time-reversal invariance in the ground
state. In addition, we restrict the Hilbert space to the SU($3N$) representation
given by the Young tableau of $N$ columns and 1 row by
imposing the constraint
$\sum_{\alpha} a^{\dagger}_{i, \alpha} a_{i, \alpha} = N$ 
(see Ref. \onlinecite{Read90PRB}).
For the following analysis we also allow for the uniform condensation 
of one of the boson species $a_{i, 1} = \sqrt{N} x$ 
to study the nematic-dimer phase transition.
The quantity $|x|^2$ then describes the nematic ``superfluid'' fraction.

\subsection{Phase diagram}
\begin{figure*}[tt]
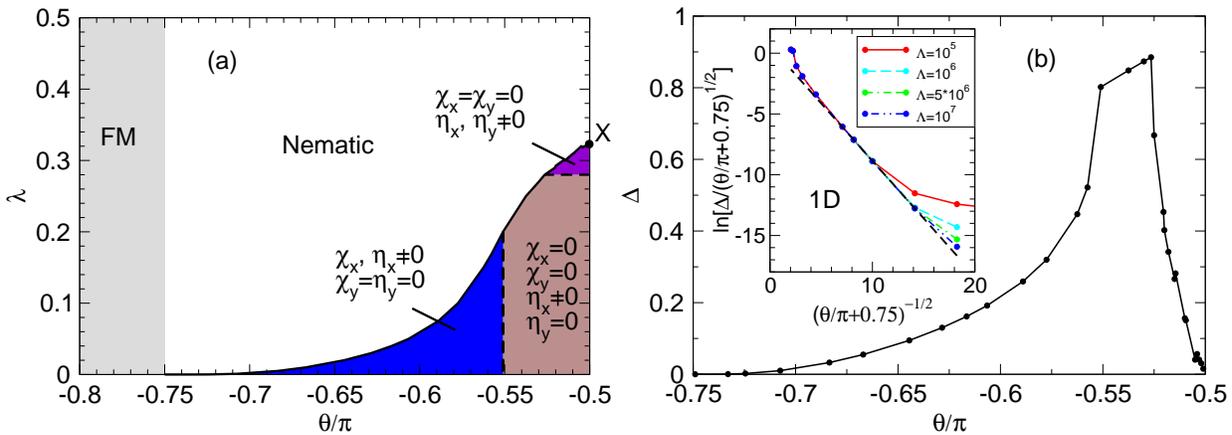

  \includegraphics*[angle=0, width=0.45\linewidth, clip]{PhaseDiagramBoldColorsV2.eps}
  \includegraphics*[angle=0, width=0.45\linewidth, clip]{EnergyGap.eps}
  \caption{(Color online) (a) Phase diagram of the anisotropic 
    bilinear bi-quadratic
    spin-1 Heisenberg model in the regime  
    $-0.75 \pi \leq \theta \leq -0.5 \pi$ 
    ($J \leq K \leq 0$) based on a Schwinger boson mean-field approach. 
    The solid line represents a first-order phase boundary except at
    $\theta = -0.5 \pi$,
    while the dashed lines represent second-order phase boundaries.
    The coordinates of the tricritical point X
    separating the nematic, the N\'{e}el ($\theta > -0.5 \pi$), 
    and the disordered phases 
    are $\theta_{\text{X}} = -0.5 \pi$ and $\lambda_{\text{X}} \approx 0.323$.
    (b) Energy gap $\Delta$ (in units of 1)
    \emph{along the spin-nematic-disorder phase boundary},
    which vanishes at $\theta = -0.5 \pi$ and $-0.75 \pi$.
    The inset shows the energy gap 
    for a one-dimensional ($\lambda = 0$) spin chain
    of various lengths $\Lambda$ near the critical point 
    at $\theta = -0.75 \pi$.
    A fit to the largest spin chain (black dashed curve) gives
    $b=\text{ln}(b')=-0.945$ and $m=0.945$
    (see main text).
    \label{fig:PhaseDiagramAndEnergyGap}}
\end{figure*}

Diagonalizing and minimizing the free energy, we 
obtain the phase diagram for the $JK$ model
in the range $-0.75 \pi \leq \theta \leq -0.5 \pi$
as shown in Fig. \ref{fig:PhaseDiagramAndEnergyGap}(a). 
For sufficiently large $\lambda$ the ground state is always a spin nematic.
However, as one tunes $\lambda$ towards smaller values 
the nematic gives way to disordered spin liquid phases
via a first order transition except at the tricritical 
points at $\theta= -0.5 \pi$ and $-0.75 \pi$, 
where the transition becomes a continuous one.
This behavior is corroborated by the 
energy gap along the phase boundary 
[Fig. \ref{fig:PhaseDiagramAndEnergyGap}(b)]. 
The gap vanishes as one approaches the points 
with enlarged SU(3) symmetry at either end of the 
nematic-disorder phase boundary at $\theta = -0.75 \pi$ and $-0.5 \pi$.
In one dimension [see inset in Fig. \ref{fig:PhaseDiagramAndEnergyGap} (b)], 
the behavior of the gap near $\theta=-0.75 \pi$ can be described by
\begin{equation}
  \label{eq:gap}
  \Delta = b' \sqrt{\theta/\pi+0.75} \, \text{exp}\Bigg(-\frac{m}
	 {\sqrt{\theta/\pi+0.75}}\Bigg),
\end{equation}
where $b'$ and $m$ are positive constants.
It is interesting to note that this form is reminiscent of the 
Berezinski-Kosterlitz-Thouless-type transition in
the (1+1)-dimensional sine-Gordon model. \cite{Minnhagen87RMP}
Equation (\ref{eq:gap}) can be obtained from the constraint equation
$\langle \Lambda^{-1} \sum_{k_{x}, \alpha} a^{\dagger}_{k_x, \alpha} a_{k_{x}, 
  \alpha} \rangle = N$
by considering the most singular contribution in the
$k_{x}=0$ and $\theta = -0.75 \pi$ limit and 
by exploiting the fact that the fields approach a common value
$\eta_{x} = \chi_{x} = 1/3$ near $\theta = -0.75 \pi$.
Here, $\Lambda$ denotes the length of the spin chain.

Our results also show no indication of a gapped nematic phase 
near the ferromagnetic phase in the one-dimensional
regime as was proposed by Chubukov. 
In fact, it is always possible to satisfy the occupancy constraint
without a nematic condensate in one dimension.
However, at the critical point 
at $\theta=-0.75\pi$ and $\lambda=0$,
the nematic phase merges from the finite $\lambda$ regime,
as shown in Fig. \ref{fig:PhaseDiagramAndEnergyGap},
so that the nematic correlation function diverges at this point.

\begin{figure}[b]
  \includegraphics*[angle=0, width=1.0\linewidth, clip]{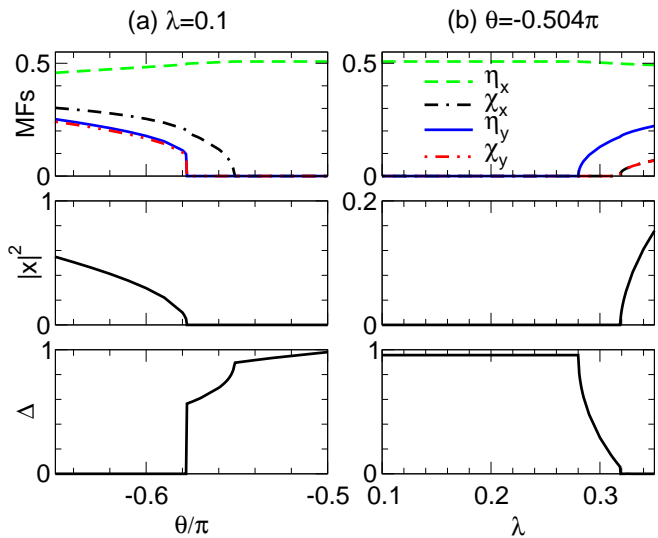}
  \caption{(Color online) Mean field values 
    $\eta_{x}, \chi_{x}, \eta_{y}, \chi_{y}$, 
    ``superfluid fraction'' $|x|^{2}$, 
    and energy gap $\Delta$ (in units of 1) for cuts (a) parallel and 
    (b) perpendicular to the $\theta$ axis of the phase diagram 
    [Fig.\ref{fig:PhaseDiagramAndEnergyGap} (a)].
    \label{fig:Cuts}}
\end{figure}

Inside the disordered regime we identify three different spin-liquid 
phases [see Fig. \ref{fig:PhaseDiagramAndEnergyGap}(a)]: 
a one-dimensional Z$_{2}$ spin-liquid ($\chi_{x}, \eta_{x} \neq 0, 
\chi_{y} = \eta_{y} = 0$), 
a one-dimensional U(1) spin-liquid ($\eta_{x} \neq 0, 
\chi_{x} = \chi_{y} = \eta_{y} = 0$),
and a two-dimensional U(1) spin-liquid ($\eta_{x}, \eta_{y} \neq 0, 
\chi_{x} = \chi_{y} = 0$).
All three phases are separated from each other by second-order 
phase boundaries that are located at $\theta_{c} \approx -0.551 \pi$ 
and $\lambda_{c} \approx 0.28$, respectively.
The decoupling of the spin chains below $\lambda_{c}$ is 
likely to be due to the mean-field scheme, which overstates the anisotropy 
in the quasi-one-dimensional limit.
Moreover, due to the one-dimensional nature, 
we expect only a weak $\lambda$-dependence of 
the phase boundary between the Z$_{2}$ and the one-dimensional U(1) phases,
which could not be resolved. 
The phase boundary between the one-dimensional and
the two-dimensional U(1) spin-liquids at $\lambda_{c}$ also does not
exhibit any dependence on the $JK$ angle $\theta$.
Detailed results along two directions in the phase diagram 
are shown in Fig. \ref{fig:Cuts}.
Last, it is interesting to note that the ``unconventional''
transition point ($\theta = -0.55 \pi$) examined previously 
in the QMC studies \cite{Harada07JPSJ}
lies in very close proximity to the Z$_{2}$-U(1) phase boundary.

Going beyond mean-field level,
saddle-point fluctuations are not likely to change the nature of the
first-order phase boundary.
We expect on the other hand that Berry phase effects will lead to 
spontaneous dimerization throughout the 
$\chi_{y}=\eta_{y}=0$ disordered region,
which lies beyond mean-field level.
This situation is similar to SU($N$) antiferromagnets,
where Berry phase effects arise from nontrivial U(1)
gauge-field fluctuations
and induce dimer ordering in 1D and 2D.
\cite{Read90PRB,Read89PRL} 
In particular, following such 
arguments for the SU(3) ($J=0$) point in the $JK$ model, we expect
a columnar dimer ordering for $\lambda < \lambda_{\text{X}}$
[see Fig. \ref{fig:PhaseDiagramAndEnergyGap}(a)]. 
We also expect that the one-dimensional $Z_{2}$
spin-liquid phase with $\chi_x,\eta_x\neq0$  is unstable towards
dimer ordering, which can be understood 
via a mapping to the odd Ising gauge theory in 1D.
\cite{Moessner01PRB}
Note that a two-dimensional $Z_{2}$ spin liquid phase
is absent on mean-field level even at finite $\lambda$. 
It is possible that small modifications to this model may reveal
a two-dimensional $Z_{2}$ spin liquid that, depending on the vison fugacity,
may or may not be stable towards dimer ordering.
\cite{Senthil00PRB}  
This is beyond the present study and will be addressed in 
the near future.


\section{Application to optical lattices}
Spin systems with a large bi-quadratic 
contribution are rarely found in solid state systems.
However, models with higher-order spin interactions
can almost perfectly be realized
with ultracold spinor atoms in optical lattices.
Moreover, anisotropy-tuned phase transitions
between different spin states
can easily be induced by
changing the optical lattice potential 
in a particular direction,
hence favoring or disfavoring exchange processes
between spins along this direction.
The spin-nematic-dimer transition also conserves magnetization, which 
is a fundamental constraint on optical lattice experiments.

Considering the lower $S=1$ hyperfine energy manifold
of $^{23}$Na atoms, a large bi-quadratic spin interaction 
naturally arises in an optical square lattice 
with a single average occupancy per site.
\cite{Imambekov03PRA,Yip03PRL}   
This follows from the two-dimensional spin-1 Bose-Hubbard model, 
\begin{eqnarray}
  H_{BH} &=& -t \sum_{\langle i, j \rangle, \sigma} 
  (b^{\dagger}_{i, \sigma} b_{j, \sigma} 
  + b^{\dagger}_{j, \sigma} b_{i, \sigma}) 
  + \frac{U_{0}}{2} \sum_{i} n_{i}(n_{i} - 1) \nonumber \\
  & &+ \frac{U_{2}}{2} \sum_{i} ({\bf S^{2}_{i}} - 2 n_{i})
  - \mu \sum_{i} n_{i}, 
\end{eqnarray}
where $b^{\dagger}_{i, \sigma}$ creates a boson on site $i$
with spin $\sigma=0, \pm1$, 
$n_{i} = \sum_{\sigma} b^{\dagger}_{i, \sigma} b_{i, \sigma}$
is the number operator, and ${\bf S}_{i}$ denotes the total spin on site $i$.
The spin-dependent potential $U_{2}$
originates from the difference in 
the scattering lengths for the two possible scattering 
spin channels $S=0$ and $2$ for
spin-1 atoms.
To leading order in $t/U_{0}$, 
the effective spin model then results in Eq.
(\ref{eq:JKModel}) (with $\lambda = 1$), where 
the exchange integrals are determined by 
$J = - 2 t^{2}/(U_{0} + U_{2})$ and
$K = - 2 t^{2}/(3 (U_{0}+ U_{2})) - 4 t^{2}/(3 (U_{0} - 2 U_{2}))$.
Typically, for $^{23}$Na atoms one has $U_{2}/U_{0} \approx 0.04$,
which translates to $J/K \approx 0.34$, or equivalently, 
$\theta \approx -0.604 \pi$ 
in the effective spin model.
In an anisotropic lattice, the hopping 
amplitude $t$ becomes directionally dependent,
giving rise to anisotropic $J$ and $K$ integrals.

How can one probe the nematic to dimer phase transition?
A common way of analyzing optical 
lattice experiments is to release 
all atoms from the trap and to measure the
column density of the expanding cloud.
Within this approach, a straightforward way
to observe a spin nematic to dimer phase transition  
\cite{Imambekov03PRA}
is to apply a weak magnetic field
(say along the 3-axis) to align the nematic hard axis 
perpendicular to the field prior to the release,
and to separate different spin components in the expanding cloud spatially 
by applying a gradient field after the release.
A sudden change in the population of the $a^{\dagger}_{3}$ state
when tuning $\lambda$ then signals the transition to the dimerized VBS.

\begin{figure}[tt]
  \includegraphics*[angle=0, width=0.9\linewidth, clip]{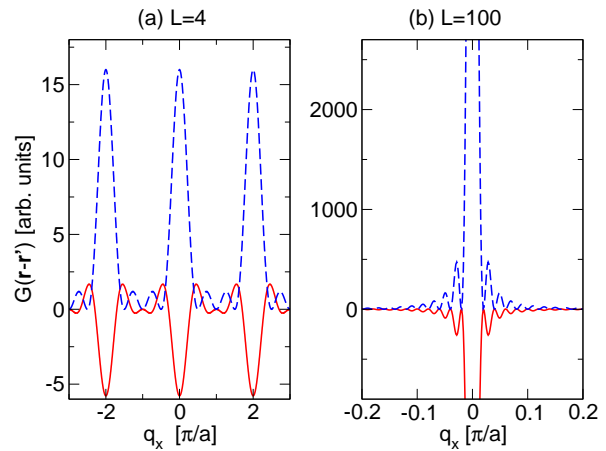}
  \caption{(Color online) Second-order density correlation function for
    the spin-nematic (blue, dashed curves) 
    and the dimer (red, solid curves) phases
    for different lattice sizes 
    along the $q_{\text{y}}=0$ direction  
    [${\bf q} = m ({\bf r} - {\bf r}')/\hbar t$].
    The central peak heights are roughly given by $\sim (1-c) L^{2}$
    with $c=2$ in the spin nematic and $4/9$
    in the dimer phase.
    Note the different $x$-axis scalings.
    \label{fig:correlation}}
\end{figure}

Besides this, the interference pattern and 
the spatial noise correlations of the expanding cloud
also provide information about the quantum state in the lattice.
\cite{Grondalski99OptEx,Altman04PRA,Roth03PRA,Foelling05Nature}
For instance, in the case of spin-1 atoms (with mass $m$),
the equal-time density-density correlation function
for the freely expanding gas after long times takes the form 
\cite{Brennen07NJP}
\begin{eqnarray}
  \label{eq:correlation}
  G({\bf r}, {\bf r}') &=& \frac{1}{L^2}\sum_{\alpha, \beta} 
  \langle n_{\alpha}({\bf r}) n_{\beta}({\bf r'})\rangle \\
  &\sim& - \frac{\text{sin}^{2}\big(q_{x} L/2 \big) 
             \text{sin}^{2}\big(q_{y} L/2 \big) }
            {L^2 W^4 \text{sin}^{2}\big(q_{x}/2 \big) 
             \text{sin}^{2}\big(q_{y}/2 \big) } \nonumber \\ 
  && + \frac{1}{L^2 W^4} \sum_{i, j} 
	    \text{e}^{i {\bf q}({\bf R}_{i}-{\bf R}_{j})}
  \langle {\bf S}_{i}{\bf S}_{j} + \big({\bf S}_{i}{\bf S}_{j} \big)^2 
  \rangle, \nonumber 
\end{eqnarray}
where ${\bf q} = m ({\bf r}-{\bf r'})/\hbar t$, 
$L$ denotes the linear size of the square lattice
in units of the lattice parameter, 
and $W$ is the width of the expanding Wannier 
states originally centered at lattice sites ${\bf R}_{i}$.
Here, we have omitted a delta term from normal ordering 
and constants of order $1/L^2$.
The first term stems from the unit occupancy constraint per site, 
while the second term contains the SU(3) spin structure factor
which constitutes the $JK$ model exactly at the 
nematic-ferromagnetic phase 
boundary ($J=K < 0 \Leftrightarrow \theta =-0.75 \pi$).
The signature of a ferromagnetic condensate therefore 
is indistinguishable from that of a spin-nematic one.
Our interest, however, lies in the nematic-dimer phase transition.
To evaluate $G({\bf r}, {\bf r}')$,  we assume
complete condensation in the $\alpha$ state 
$|\phi_\text{{nem}}\rangle = \prod_{i} a^{\dagger}_{i, \alpha} |0\rangle$
in the spin nematic phase and obtain $\langle {\bf S}_{i}{\bf S}_{j} 
+ \big({\bf S}_{i}{\bf S}_{j} \big)^2\rangle_{\text{nem}} = 2$
for all $i, j$. 
In the dimer phase one gets $\langle {\bf S}_{i}{\bf S}_{j} 
+ \big({\bf S}_{i}{\bf S}_{j} \big)^2 \rangle_{\text{dimer}} = 2$ if the spins 
on sites $i$ and $j$ form a singlet and $4/9$ otherwise.
The correlation function then takes the form
\begin{eqnarray}
  G({\bf r}, {\bf r}') &\sim& - \frac{(1-c) \text{sin}^{2}\big(q_{x} L/2 \big) 
    \text{sin}^{2}\big(q_{y} L/2 \big) }
  {L^2 W^4 \text{sin}^{2}\big(q_{x}/2 \big) 
    \text{sin}^{2}\big(q_{y}/2 \big) } \nonumber \\ 
  && + \frac{2-c}{W^4} \big(\text{cos}(q_{x})
  + \text{cos}(q_{y}) \big) 
\end{eqnarray}
with $c=2$ in the nematic and $4/9$ in the dimer phase.
By measuring the density density correlator,
both phases can be clearly distinguished 
as shown in Fig. \ref{fig:correlation}.
Assuming a lattice spacing of 532 nm, a time of flight 
of 20 ms and a detector size of $5 \, \mu\text{m}$,
the momentum resolution results in $\Delta q/(\pi/a) \approx 0.015$
sufficient to observe the peaks for 
an $L \approx 100$ lattice ($\text{FWHM}/(\pi/a)  \approx 0.02$, where FWHM stands for full width at half maximum).

Lastly, a recent and interesting proposal for probing the spin configuration 
in an optical lattice is based on polarization spectroscopy.
\cite{Eckert08NaturePhys,Eckert07PRL}
This kind of measurement leaves the lattice intact
as a signature of the spin state is
imprinted in the polarization of a probing light beam.
By analyzing the noise fluctuations
in the polarization of the outgoing light,
the dimerized and the spin-nematic phase can be well
discriminated.

\section{Conclusion}
In summary, we have considered the spin nematic to dimer phase transition 
in the anisotropic $JK$ model using a bond operator formalism 
and a large-$N$ mean field approach.
Our analysis generally suggests 
a first-order spin nematic to dimer transition
but would not contradict the possibility
of deconfined criticality 
at $J=0$ ($\theta = -0.5 \pi$), where the $JK$ model
has an enlarged SU(3) symmetry.
Our large-$N$ analysis 
reveals a critical point at $J=K < 0$ ($\theta = -0.75 \pi$)
where two phase-transition lines, 
separating the spin-liquid, 
spin-nematic, and ferromagnetic phases, meet.
Although the nematic phase vanishes in 1D, the nematic correlation length 
diverges at the critical point.
We have also argued that the 1D $Z_{2}$
and the 1D and 2D U(1) spin-liquids are unstable toward dimer ordering
while the stability of a 2D $Z_{2}$ spin liquid 
would depend on the vison fugacity.
We do not find a 2D $Z_{2}$ spin-liquid at mean-field level
but will address a route to such a phase in the future.
Finally, we have discussed various ways of observing  
a spin-nematic to dimer phase transition in optical lattices,
where the lattice anisotropy can be tuned through laser intensities.
Such experiments eventually provide concrete ways of 
studying the phases and phase transitions of the $JK$ model.

\begin{acknowledgments}
We thank A.~V. Chubukov,  D. Podolsky, S. Sachdev, T. Senthil, M. Troyer 
and J. Thywissen 
for valuable discussions. 
H.Y.K. thanks KITP for hospitality where this work was initiated. 
This work was supported 
by NSERC of Canada, Canada Research
Chair and the Canadian Institute for Advanced Research.
\end{acknowledgments}

\end{document}